\documentclass[pra,a4paper,twoside,twocolumn,showpacs]{revtex4}
\usepackage{amsmath,amsfonts,graphicx,mathptmx}
\usepackage[amssymb]{SIunits}

\newcommand{\setC}{\mathbb{C}}
\newcommand{\ket}[1]{|#1\rangle}
\newcommand{\bra}[1]{\langle#1|}
\newcommand{\bracket}[2]{\langle#1|#2\rangle}
\DeclareMathOperator{\tr}{tr}

\DeclareMathOperator{\Choi}{Choi}
\SetSymbolFont{symbols}{normal}{OMS}{cmsy}{m}{n}

\begin{document}
\bibliographystyle{apsrev}
\title{Unstable particles as open quantum systems}

%\author{Pawe{\l}{} Caban \and Jakub Rembieli{\'n}ski \and Kordian A. Smoli{\'n}ski
%  \and Zbigniew Walczak}
\author{Pawe{\l}{} Caban}
\email{P.Caban@merlin.fic.uni.lodz.pl}
\author{Jakub Rembieli{\'n}ski} 
\email{J.Rembielinski@merlin.fic.uni.lodz.pl}
\author{Kordian A. Smoli{\'n}ski}
\email{K.A.Smolinski@merlin.fic.uni.lodz.pl}
\author{Zbigniew Walczak}
\email{Z.Walczak@merlin.fic.uni.lodz.pl}

\affiliation{Department of Theoretical Physics, University of Lodz\\
Pomorska 149/153, 90-236 {\L}{\'o}d{\'z}, Poland}

\date{14 September 2005}
\begin{abstract}
  We present the probability-preserving description of the decaying
  particle within the framework of quantum mechanics of open systems,
  taking into account the superselection rule prohibiting the
  superposition of the particle and vacuum.  In our approach the
  evolution of the system is given by a family of completely positive
  trace-preserving maps forming a one-parameter dynamical semigroup.
  We give the Kraus representation for the general evolution of such
  systems, which allows one to write the evolution for systems with
  two or more particles.  Moreover, we show that the decay of the
  particle can be regarded as a Markov process by finding explicitly
  the master equation in the Lindblad form.  We also show that there
  are remarkable restrictions on the possible strength of decoherence.
\end{abstract}
\maketitle

\section{Introduction}
\label{sec:introduction}

Recently, the tests of Bell inequalities \cite{bell65} in the system
of correlated neutral kaons \cite{bertlmann01a,quant-ph/0410028} or
$B$ mesons \cite{go04} has attracted some attention.  The crucial
point in studying correlations in this system are the oscillations of
the strangeness and bottom, respectively.  However, the instability of
kaons makes the analysis of correlation experiments difficult.  The
state of the complete system is a superposition (or a mixture) of the
states of the decaying particles and the decay products.  The whole
system undergoes a unitary evolution described usually in terms of
quantum field theory.  On the other hand, in correlation experiments
of Einstein--Podolsky--Rosen--Bohm type \cite{einstein35,bohm51}, it
is more useful to neglect the evolution of decay products and consider
solely the decaying particles.  Unfortunately, such a description
within the framework of quantum mechanics referring to the case with
finite degrees of freedom leads to some difficulties.  This is usually
done by means of the Weisskopf--Wigner approach
\cite{weisskopf30a,weisskopf30b}, where the probability of detecting
the particle is not conserved during the time evolution and,
therefore, the Hamiltonian in such theories must be non-Hermitian.
Moreover, this formalism leads to some ambiguities when applied to the
description of correlation experiments.  The reason is the probability
loss caused by the decrease of the trace of the reduced density
operator.  This prevents one to calculate unambiguously the
probability of finding the system in a given state after the
projective measurement.  Therefore, in our opinion, we need an
approach enabling a description of the system that can be in
two-particle states as well as the one-particle and even zero-particle
states, which can arise during the time evolution (decay) of the
initial system.  It seems to us that the mentioned issues can be
resolved by an assumption that the decaying particle can be found in a
particle state as well as in the state of \emph{the absence of the
  particle}, i.e., in the \emph{vacuum} state (it is not a vacuum in
the sense used in quantum field theory, but rather in a sense used in
\cite{caban02}).

In this paper, we give the probability preserving description of the
decaying particle within the framework of quantum mechanics of open
systems.  This approach is introduced in Sec.~\ref{sec:evolution},
where the evolution of the system is given by a family of completely
positive trace preserving maps forming a one-parameter dynamical
semigroup \cite{kossakowski72,gorini76,lindblad76}.  Thus, in our
approach the Hamiltonian is Hermitian and therefore the reduced
density operator has a unit trace.  We also find the operator-sum
representation (Kraus representation) for the evolution of such
systems, which immediately allows one to write the evolution for
systems of two or more noninteracting particles.  This is useful if
we study quantum correlations between unstable particles.  We would
like to point out that we use the dynamical semigroup approach for the
entire evolution of the unstable particle, not only for the
description of its decoherence, as was done in
\cite{benatti96,benatti97a,benatti97b,benatti98,quant-ph/0410028}.
%In the Sec.~\ref{sec:prob-aver-state} we give a few examples of direct
%applications of our formalism.
Finally, in Sec.~\ref{sec:cp-viol-decoh} we study the restrictions on
the possible strength of decoherence that arise as a side effect of
completely positive evolution of the system as well as we estimate the
upper bound for the decoherence strength for $K^0$ and $B^0$ mesons.

\section{The time evolution of unstable particles}
\label{sec:evolution}

In this section we discuss the evolution of unstable particles,
neglecting their spatial degrees of freedom.  In order to sketch our
approach, we begin with the discussion of the case of the neutral
pion.  This example is rather elementary, but it helps us to
illustrate the main idea of our approach.  Next, we go to $K^0$
($B^0$) mesons.  The evolution of these particles is more complicated
because of the phenomenon of transmutation between $K^0$ and
$\bar{K^0}$ ($B^0$ and $\bar{B}^0$).  We shall regard them as open
systems, it means that their evolution is not unitary, but it must be
treated as a one-parameter family of quantum operations forming a
dynamical semigroup.  Consequently, the density operator of the system
must obey the master equation rather than the von Neumann equation.

\boldmath
\subsection{Unstable (pseudo)scalar particle}
\label{sec:pion}
\unboldmath

In this section we describe briefly the evolution of an unstable
(pseudo)scalar particle, $\pi^0$, and we show that this evolution can be
regarded as a family of amplitude damping quantum operations
\cite{nielsen00}.

The key point of the presented approach is that the system under
consideration can be regarded as a two-level system: one can find the
system in the particle state or in the vacuum state.  Of course, this
system must be an open one and we treat the decay products as a part
of the environment.

The space of states of the system is a direct sum of the Hilbert space
of the particle $\mathcal{H}_{\pi^0}$, spanned by the vector
$\ket{\pi^0}$, and the Hilbert space of the vacuum $\mathcal{H}_{0}$,
spanned by the vector $\ket{0}$; i.e., $\mathcal{H} =
\mathcal{H}_{\pi^0} \oplus \mathcal{H}_{0}$.  %Moreover, the vacuum $\ket{0}$
%is orthogonal to any vector from $\mathcal{H}_{\pi^0}$.
Since we are looking for a description of a decaying particle in terms
of quantum mechanics with finite degrees of freedom (not in the
language of field theory; cf.\ \cite{alicki78}) we assume that the
decay process is a Markov process and it is not described in the
dynamical manner (i.e., it is not governed by a Hamiltonian).
%Therefore the particle number operator
%generates a superselection rule, allowing the observable to be build
%from the projections either on the particle state $\ket{\pi^0}$ or on
%the vacuum $\ket{0}$ but it cannot contain projections on the
%superpositions of the particle and the vacuum.  Equivalently, we can
%restrict the admissible set of physical states to ones with density
%operator commuting with the particle number operator $\hat{N} =
%\ket{\pi^0} \bra{\pi^0}$.
%Because the
%particle usually carries some conserved quantum numbers we have to
%impose a superselection rule prohibiting superpositions of $\ket{\pi^0}$
%and $\ket{0}$.  Therefore, we are looking for the evolution preserving
%the superselection rule, i.e., if the density operator obeys the
%superselection rule at time $t = 0$, it obeys this rule for any time
%$t \geq 0$.

We represent the vectors $\ket{\pi^0}$ and $\ket{0}$ by
\begin{equation}
  \label{eq:basis-1-pion}
  \ket{\pi^0} =
  \begin{pmatrix}
    1 \\ 0
  \end{pmatrix}, \quad
  \ket{0} =
  \begin{pmatrix}
    0 \\ 1
  \end{pmatrix}.
\end{equation}

The time evolution of the system can be represented by the continuous
one-parameter family of linear superoperators $\mathcal{S}_t$ such
that
\begin{equation}
  \label{eq:evolution}
  \hat{\rho}(t) = \mathcal{S}_t \hat{\rho}(0),
\end{equation}
where $\hat{\rho}(t)$ is the density operator of the system at the time
$t$.  These superoperators must be trace-preserving completely
positive maps and they must form a one-parameter semigroup (see, e.g.,
\cite{alicki01,breuer02} and refereces therein), i.e.,
\begin{subequations}
  \label{eq:semi-group}
  \begin{gather}
%    \mathcal{S}_0 = \id,\\
    \tr[\mathcal{S}_t \hat{\rho}(0)] = \tr[\hat{\rho}(0)] = 1,\\
    \label{eq:comp-law}
    \mathcal{S}_{t_1 + t_2} = \mathcal{S}_{t_1}
    \mathcal{S}_{t_2},\quad \forall t_1, t_2 \geq 0,
  \end{gather}
\end{subequations}
and the map $t \mapsto \mathcal{S}_t$ is continuous in strong topology.
%Moreover, if we impose the superselection rule on the initial state,
%the states of the system at any time must be subjected to this
%superselection rule.

We study the time evolution of the state of the system,
\begin{equation}
  \label{eq:state}
  \hat{\rho} = \rho_{11} \ket{\pi^0} \bra{\pi^0} + \rho_{12} \ket{\pi^0} \bra{0} 
  + \rho_{12}^* \ket{0} \bra{\pi^0} + \rho_{22} \ket{0} \bra{0},
\end{equation}
assuming that it is consistent with phenomenological Weisskopf--Wigner
evolution \cite{weisskopf30a,weisskopf30b},
\begin{equation}
  \label{eq:WW-pi}
  \ket{\pi^0(t)}  = e^{-t (i m + \Gamma/2)} \ket{\pi^0},
\end{equation}
where $m$ is the $\pi^0$ mass and $\Gamma$ is its decay width.

From~\eqref{eq:WW-pi}, it follows that $\rho_{11}(t) = e^{-t \Gamma} \rho_{11}(0)$
and therefore $\rho_{22}(t) = 1 - e^{-t \Gamma} \rho_{11}(0)$.  Taking into
account the linearity of $\mathcal{S}_t$, we can write $\rho_{12}(t)$ as
the time-dependent linear combinations of all the elements of the
initial density matrix, i.e.\ $\rho_{12}(t) = \sum_{i,j=1}^2 A_{ij}(t)
\rho_{ij}(0)$, with the initial conditions $A_{12}(0) = 1$ and all
remaining $A$'s vanish at $t = 0$.

Therefore the action of the map $\mathcal{S}_t$ can be written as
follows:
\begin{equation}
  \label{eq:S-rho:pi}
  \mathcal{S}_t \rho(0) = \rho(t) \\ 
  =
  \begin{pmatrix}
    e^{-t \Gamma} \rho_{11}(0) & \sum\limits_{i,j=1}^2 A_{ij}(t) \rho_{ij}(0) \\
    \sum\limits_{i,j=1}^2 A_{ij}^*(t) \rho_{ij}^*(0) & \rho_{22}(t)
    % \rho_{22}(0) + \left(1 - e^{-t \Gamma}\right) \rho_{11}(0)
  \end{pmatrix}
  ,
\end{equation}
where
\begin{equation}
  \label{eq:S-rho00:pi}
  \tag{\ref{eq:S-rho:pi}a}
  \rho_{22}(t) = \rho_{22}(0) + \left(1 - e^{-t \Gamma}\right) \rho_{11}(0).
\end{equation}

To find conditions under which the map \eqref{eq:S-rho:pi} is
completely positive, we use the Choi's theorem \cite{choi75,alicki01},
which states that $\mathcal{S}_t$ is completely positive iff the
corresponding Choi's matrix, %(see \cite{havel03})
\begin{equation}
  \label{eq:choi-pi}
  \Choi\mathcal{S}_t =
  \begin{pmatrix}
    e^{-t \Gamma} & A_{11}(t) & 0 & A_{12}(t) \\
    A_{11}^*(t) & 1 - e^{-t \Gamma} & A_{21}^*(t) & 0 \\
    0 & A_{21}(t) & 0 & A_{22}(t) \\
    A_{12}^*(t) & 0 & A_{22}^*(t) & 1
  \end{pmatrix}
\end{equation}
is positive.  This implies that
\begin{subequations}
  \allowdisplaybreaks
  \label{eq:positive}
  \begin{align}
    \label{eq:positiv12}
    |A_{12}(t)|^2 &\leq e^{-t \Gamma},\\
    \label{eq:positiv11}
    |A_{11}(t)|^2 &\leq \left(1 - e^{-t \Gamma}\right) \left(e^{-t \Gamma} - |A_{12}(t)|^2\right),\\
    A_{21}(t) &= 0,\\
    A_{22}(t) &= 0.
  \end{align}
\end{subequations}

One can check by straightforward calculation that the composition law
(\ref{eq:comp-law}) leads to the conditions
\begin{subequations}
  \allowdisplaybreaks
  \label{eq:compositon}
  \begin{gather}
    \label{eq:a12}
    A_{12}(t_1 + t_2) = A_{12}(t_2) A_{12}(t_1),\\
    \label{eq:a11}
    A_{11}(t_1 + t_2) = A_{11}(t_2) e^{-t_1 \Gamma} 
    +A_{12}(t_2) A_{11}(t_1).
  \end{gather}
\end{subequations}
It is easy to see that the only possible solution of~\eqref{eq:a12}
fulfilling~\eqref{eq:positiv12} and the initial conditions is
\begin{equation}
  \label{eq:A12}
  A_{12}(t) = e^{-t \left[(\Gamma + \lambda)/2 + i \mu\right]},
\end{equation}
where $\lambda \geq 0$ and $\mu \geq 0$.  Taking into account the fact that the
one-parameter semigroup must be Abelian, we get from~(\ref{eq:a11}),
and~(\ref{eq:A12})
\begin{multline}
  \label{eq:A11}
  A_{11}(t) = \\
  \begin{cases}
    z \left\{e^{-t \Gamma} - e^{-t \left[(\Gamma + \lambda)/2 + i \mu\right]}\right\}, & \text{when } \lambda \neq \Gamma
    \text{ or } \mu \neq 0,\\
    z t e^{-t \Gamma}, & \text{when } \lambda = \Gamma \text{ and } \mu = 0,
  \end{cases}
\end{multline}
where $z \in \setC$ are such that the inequality~(\ref{eq:positiv11}) is
satisfied (for $\lambda = 0$ we have to put $z = 0$).  Therefore, the most
general form of the time-dependent density matrix is given by
\begin{widetext} 
\begin{equation}
  \label{eq:matrix}
  \rho(t) =
  \begin{pmatrix}
    e^{-t \Gamma} \rho_{11}(0) & e^{- t \left[(\Gamma + \lambda)/2 + i m\right]} \rho_{12}(0) + A_{11}(t) \rho_{11}(0) \\
    e^{-t \left[(\Gamma + \lambda)/2 - i m\right]} \rho^*_{12}(0) + A_{11}^*(t) \rho_{11}(0) & 1 - e^{-t \Gamma} \rho_{11}(0)
  \end{pmatrix},
\end{equation}
\end{widetext}
where the consistency with (\ref{eq:WW-pi}) requires $\mu = m$.  The
parameter $\lambda$ is interpreted as the decoherence parameter.

Since the evolution of the system~\eqref{eq:matrix} is given by a
completely positive map, it can also be written in the operator-sum
form \cite{kraus83}
\begin{equation}
  \label{eq:Kraus}
  \hat{\rho}(t) = \sum_{i=0}^N \hat{E}_i(t) \hat{\rho}(0) \hat{E}_i^{\dag}(t),
\end{equation}
where the Kraus operators $\hat{E}_i(t)$ satisfy the condition
$\sum_{i=0}^N \hat{E}_i^{\dag}(t) \hat{E}_i(t) = I$.  One can easily check
that the Kraus operators leading to the evolution~\eqref{eq:matrix}
are given by
\begin{subequations}
  \allowdisplaybreaks
  \label{eq:1-pion-kraus}
  \begin{gather}
    \hat{E}_0(t) = e^{-t \left[(\Gamma + \lambda)/2 + i m\right]} \ket{\pi^0} \bra{\pi^0} 
    + \ket{0} \bra{0},\\
    \hat{E}_1(t) = \sqrt{1 - e^{-t \Gamma} - \frac{|A_{11}(t)|^2 e^{t \Gamma}}{1 - e^{-t \lambda}}}
    \ket{0} \bra{\pi^0},\\
    \hat{E}_2(t) = e^{-t \Gamma/2} \sqrt{1 - e^{-t \lambda}} \ket{\pi^0} \bra{\pi^0}
    + \frac{A_{11}^*(t) e^{t \Gamma/2}}{\sqrt{1 - e^{-t \lambda}}}  \ket{0} \bra{\pi^0}.
  \end{gather}
\end{subequations}

From the operator-sum representation, using standard procedures
\cite{havel03,fisher04}, we can easily find the local form of the time
evolution---the master equation in the Lindblad form \cite{lindblad76},
\begin{equation}
  \label{eq:master}
  \frac{d\hat{\rho}(t)}{dt} = -i [\hat{H}, \hat{\rho}(t)] + \{\hat{K}, \hat{\rho}(t)\}
  + \sum_{i=1}^N \hat{L}_i \hat{\rho}(t) \hat{L}_i^{\dag}, 
\end{equation}
where $\hat{H}$ is the Hamiltonian of the system, the operators
$\hat{L}_i$ are the Lindblad operators, and $\hat{K} = -\frac{1}{2}
\sum_{i=1}^N \hat{L}_i^{\dag}{} \hat{L}_i$.  For the density
operator~\eqref{eq:matrix}, the Hamiltonian is
\begin{subequations}
  \label{eq:1-pion}
  \begin{gather}
    \label{eq:1-pion-hamiltonian}
    \hat{H} = m \ket{\pi^0} \bra{\pi^0},%
    \intertext{and the Lindblad operators are of the form}
    \label{eq:1-pion-lindblad}
    \hat{L}_1 = \sqrt{\Gamma (1 - \alpha)} \ket{0} \bra{\pi^0},\\
    \hat{L}_2 = \sqrt{\lambda} \ket{\pi^0} \bra{\pi^0} 
    + \frac{\beta^*}{\sqrt{\lambda}} \ket{0} \bra{\pi^0},
  \end{gather}
\end{subequations}
where
\begin{subequations}
\begin{gather}
  \label{eq:alfa}
  \alpha = %\lim_{t \searrow 0} \frac{|A_{11}(t)|^2}{(1 - e^{-t \Gamma}) (1 - e^{-t \lambda})} =
  \begin{cases}
    |z|^2 [4 m^2 + (\Gamma - \lambda)^2]/(4 \Gamma \lambda), & \lambda \neq \Gamma \text{ or } m \neq 0,\\
    |z|^2/\Gamma^2, & \lambda = \Gamma \text{ and } m = 0,
  \end{cases}
  \\
  \beta = %A_{11}'(0) = 
  \begin{cases}
    z |i m + (\Gamma - \lambda)/2|, & \lambda \neq \Gamma \text{ or } m \neq 0,\\
    z, & \lambda = \Gamma \text{ and } m = 0.
  \end{cases}
\end{gather}
\end{subequations}

Now, we take into account the fact that superpositions of the particle
and vacuum are not observed in the nature.  Consequently, there is no
physical observable with nonvanishing matrix elements between vacuum
state and particle state, which leads to the \emph{superselection rule}.
Therefore, the element $\rho_{12}$ of the density operator
(\ref{eq:state}) does not contribute to the expectation value of any
observable, and we can assume that $\rho_{12}(t) = 0$ for any time
\cite{ballentine98}, which implies that $z = 0$.  Therefore in this
case the decoherence parameter $\lambda$ becomes irrelevant, so we are free
to put $\lambda = 0$.  Thus the density matrix describing the evolution of
$\pi^0$ is
\begin{equation}
  \label{eq:pi-evol}
  \rho(t) = 
  \begin{pmatrix}
    e^{-t \Gamma} \rho_{11}(0) & 0 \\
    0 & 1 - e^{-t \Gamma} \rho_{11}(0)
  \end{pmatrix}
  .
\end{equation}
The corresponding Kraus operators have the following form:
\begin{subequations}
  \label{eq:1-pion-kraus-ss}
  \begin{gather}
    \hat{E}_0(t) = e^{-t \left[\Gamma/2 + i m\right]} \ket{\pi^0} \bra{\pi^0} 
    + \ket{0} \bra{0},\\
    \hat{E}_1(t) = \sqrt{1 - e^{-t \Gamma}} \ket{0} \bra{\pi^0},
  \end{gather}
\end{subequations}
while the generators of the master equation are
\begin{subequations}
  \label{eq:1-pion-master-ss}
  \begin{gather}
    \hat{H} = m \ket{\pi^0} \bra{\pi^0} ,\\
    \hat{L}_1 = \sqrt{\Gamma} \ket{0} \bra{\pi^0}.
  \end{gather}
\end{subequations}

Note that the evolution of $\pi^0$ is thus simply the amplitude damping
quantum operation \cite{nielsen00} with the probability of damping
depending on time, namely $p = 1 - e^{-t \Gamma}$.  If the initial state of
the system is $\hat{\rho}(0) = \ket{\pi^0} \bra{\pi^0}$, then, as expected,
the probability of detecting $\pi^0$ at the time $t$ is
\begin{equation}
  \label{eq:Geiger-Nutall}
  p(\pi^0) = \tr[\hat{\rho}(t) \ket{\pi^0} \bra{\pi^0}] = e^{-t \Gamma},
\end{equation}
i.e., it is given by the Geiger--Nutall law. 

\boldmath
\subsection{The time evolution of $K^0$ ($B^0$)}
\label{sec:evolution-k0}
\unboldmath

Now, we consider the case of a $K^0$ ($B^0$) meson.  This particle
needs special treatment because during the time evolution it
transmutes into its antiparticle.  Because both $K^0$ and $B^0$ mesons
evolve according to the same scheme, hereafter we shall deal with
$K^0$, but the results are also valid for $B^0$ after appropriate
changes of notation.

The Hilbert space of the kaon--vacuum system $\mathcal{H}_{K^0} \oplus
\mathcal{H}_0$ is spanned by orthonormal vectors $\ket{K^0}$,
$\ket{\bar{K}^0}$, and $\ket{0}$, which are the eigenstates of the
strangeness operator $\hat{S}$:
\begin{equation}
  \label{eq:S-basis}
  \hat{S} \ket{K^0} = \ket{K^0}, \quad 
  \hat{S} \ket{\bar{K}^0} = -\ket{\bar{K}^0}, \quad
  \hat{S} \ket{0} = 0.
\end{equation}
These states (except of $\ket{0}$) are not eigenstates of the operator
$\hat{C}\hat{P}$, where $\hat{P}$ is the space reflection and
$\hat{C}$ is the charge conjugation.  The $\hat{C}\hat{P}$ eigenstates
$\ket{K^0_1}$ and $\ket{K^0_2}$,
\begin{equation}
  \label{eq:CP-basis}
  \hat{C}\hat{P} \ket{K^0_1} = \ket{K^0_1}, \quad 
  \hat{C}\hat{P} \ket{K^0_2} = -\ket{K^0_2}, \quad
  \hat{C}\hat{P} \ket{0} = \ket{0},
\end{equation}
are related to $\hat{S}$ eigenstates by
\begin{subequations}
  \label{eq:CP-S}
  \begin{gather}
    \ket{K^0_1} = \frac{1}{\sqrt{2}} \left(\ket{K^0} 
      + \ket{\bar{K}^0}\right),\\
    \ket{K^0_2} = \frac{1}{\sqrt{2}} \left(\ket{K^0} 
      - \ket{\bar{K}^0}\right).
  \end{gather}
\end{subequations}

On the other hand, the time evolution operator is not diagonal in the
basis~\eqref{eq:CP-basis} due to $CP$ violation, but it is diagonal in
the basis (see, e.g., \cite{kleinknecht03})
\begin{subequations}
  \label{eq:H-basis}
  \begin{gather}
    \ket{K^0_S} = \frac{1}{\sqrt{1 + |\epsilon|^2}} \left(\ket{K^0_1} 
      + \epsilon \ket{K^0_2}\right),\\
    \ket{K^0_L} = \frac{1}{\sqrt{1 + |\epsilon|^2}} \left(\epsilon \ket{K^0_1} 
      + \ket{K^0_2}\right),
  \end{gather}
\end{subequations}
with $\epsilon$ being the complex CP-violation parameter, $|\epsilon| \approx 2.284 \times
10^{-3}$ \cite{eidelman04:abb}.  The time evolution in this basis is
assumed to follow the Weisskopff--Wigner phenomenological prescription,
\begin{subequations}
  \label{eq:WW-K}
  \begin{gather}
    \ket{K^0_S(t)} = e^{-t (i m_S + \Gamma_S/2)} \ket{K^0_S},\\
    \ket{K^0_L(t)} = e^{-t (i m_L + \Gamma_L/2)} \ket{K^0_L},
  \end{gather}
\end{subequations}
where $\Gamma_S$ and $\Gamma_L$ are decay widths of $K^0_S$ and $K^0_L$,
respectively; $m_S$ and $m_L$ are some parameters---their physical
meaning is provided by the formulas~(\ref{eq:mSmL}).  We would like to
point out that these masses cannot be the eigenvalues of a Hermitian
Hamiltonian because of a CP violation.  Indeed, the
basis~\eqref{eq:H-basis} is no longer orthonormal, since
\begin{equation}
  \label{eq:KSKL}
  \bracket{K^0_S}{K^0_L} = \frac{2 \Re(\epsilon)}{1 + |\epsilon|^2} \equiv \delta_L \approx 3.27 \times 10^{-3},
\end{equation}
and these states \emph{cannot be the eigenstates of a Hermitian
  operator}.  % We shall return to this question after finding the
% explicit form of Hamiltonian.

The most convenient way of analyzing the evolution of the density
operator is to decompose it as follows:
\begin{align}
  \hat{\rho}(t) %\\ 
  &= 
  \tilde{\rho}_{SS} (t) \ket{K^0_S} \bra{K^0_S} + \tilde{\rho}_{SL}(t) \ket{K^0_S} \bra{K^0_L} 
  + \tilde{\rho}_{S0}(t) \ket{K^0_S} \bra{0} \notag\\
  &\quad
  + \tilde{\rho}_{SL}^*(t) \ket{K^0_L} \bra{K^0_S} + \tilde{\rho}_{LL}(t) \ket{K^0_L} \bra{K^0_L} 
  + \tilde{\rho}_{L0}(t) \ket{K^0_L} \bra{0} \notag\\
  &\quad
  + \tilde{\rho}_{S0}^*(t) \ket{0} \bra{K^0_S} + \tilde{\rho}_{L0}^*(t) \ket{0} \ket{K^0_L}
  + \tilde{\rho}_{00}(t) \ket{0} \bra{0}.
  \label{eq:rho-rho}
\end{align}
The superselection rule for a $K^0$ meson allows one to put
$\tilde{\rho}_{S0}(t) = \tilde{\rho}_{L0}(t) = 0$ for physical states.

Because the basis \eqref{eq:H-basis} is nonorthogonal, the matrix
$\tilde{\rho}(t)$ built from the coefficients of the
decomposition~\eqref{eq:rho-rho} \emph{is not formed from matrix
  elements of the density operator} $\hat{\rho}(t)$ in this basis,
therefore one should be careful while operating on $\tilde{\rho}(t)$,
especially $\tr[\hat{\rho}(t)] = 1$ implies that
\begin{equation}
  \label{eq:tr-rho}
  \tr[\tilde{\rho}(t)] = 1 - 2 \delta_L \Re[\tilde{\rho}_{SL}(t)].
\end{equation}
(see Appendix~\ref{sec:oper-sum-repr}).

Let us denote by $\rho(t)$ the matrix formed by matrix elements of
$\hat{\rho}(t)$ in the orthonormal basis $\{\ket{K^0_1}, \ket{K^0_2},
\ket{0}\}$.  From~\eqref{eq:H-basis} it follows that these two matrix
representations of $\hat{\rho}(t)$ are connected by
\begin{equation}
  \label{eq:VrhoV}
  \rho(t) = V \tilde{\rho}(t) V^{\dag},
\end{equation}
where
\begin{equation}
  \label{eq:V-mat}
  V = 
  \begin{pmatrix}
    (1 + |\epsilon|^2)^{-1/2} & \epsilon (1 + |\epsilon|^2)^{-1/2} & 0 \\
    \epsilon (1 + |\epsilon|^2)^{-1/2} & (1 + |\epsilon|^2)^{-1/2} & 0 \\
    0 & 0 & 1
  \end{pmatrix}
  .
\end{equation}

Now, we find the time evolution of the density operator in terms of
the matrix $\tilde{\rho}$.  Using~\eqref{eq:VrhoV} we can write
\begin{equation}
  \label{eq:S-rho-12}
  \rho(t) = \mathcal{S}_t \rho(0) = \mathcal{S}_t V \tilde{\rho}(0) V^{\dag}
  = V \tilde{\mathcal{S}}_t \tilde{\rho}(0) V^{\dag},
\end{equation}
where $\tilde{\rho}(t) = \tilde{\mathcal{S}}_t \tilde{\rho}(0)$
[cf.~\eqref{eq:VrhoV}].

The maps $\mathcal{S}_t$ must be completely positive and must form a
one-parameter semigroup.  By virtue of~\eqref{eq:S-rho-12} the maps
$\mathcal{S}_t$ are composed from $\tilde{\mathcal{S}}_t$ and the
map~\eqref{eq:VrhoV}.  One can easily check using Choi's theorem that
the map~\eqref{eq:VrhoV} is completely positive, so the maps
$\mathcal{S}_t$ are completely positive iff the maps
$\tilde{\mathcal{S}}_t$ are also completely positive.  It is much
easier to find the conditions under which the latter maps are
completely positive.

%After checking this conditions for maps preserving the superselection
%rule (see App.~\ref{sec:time-depend}), we get the following time
%evolution of the matrix $\tilde{\rho}(t)$:
%\begin{widetext}
%\begin{equation}
%  \label{eq:rho-kaon-unph}
%  \tilde{\rho}(t) = 
%  \begin{pmatrix}
%    e^{-t \Gamma_S} \tilde{\rho}_{SS}(0) & e^{-t (\Gamma + \lambda - i \Delta m)} \tilde{\rho}_{SL}(0) &
%    e^{-t (\Gamma_S/2 + \lambda + i m_S)} \tilde{\rho}_{S0}(0) \\
%    e^{-t (\Gamma + \lambda + i \Delta m)} \tilde{\rho}^*_{SL}(0) & e^{-t \Gamma_L} \tilde{\rho}_{LL}(0) & 
%    e^{-t (\Gamma_L/2 + \lambda + i m_L)} \tilde{\rho}_{L0}(0) \\
%    e^{-t (\Gamma_S/2 + \lambda - i m_S)} \tilde{\rho}_{S0}^*(0) 
%    & e^{-t (\Gamma_L/2 + \lambda - i m_L)} \tilde{\rho}_{L0}^*(0) & \tilde{\rho}_{00}(t)
%  \end{pmatrix}
%\end{equation}

After checking the conditions for complete positivity (see
Appendix~\ref{sec:time-depend}) and taking into account the
superselection rule, we get the following time evolution of the matrix
$\tilde{\rho}(t)$:
\begin{equation}
  \label{eq:rho-kaon}
  \tilde{\rho}(t) = 
  \begin{pmatrix}
    e^{-t \Gamma_S} \tilde{\rho}_{SS}(0) & e^{-t (\Gamma + \lambda - i \Delta m)} \tilde{\rho}_{SL}(0) & 0 \\
    e^{-t (\Gamma + \lambda + i \Delta m)} \tilde{\rho}^*_{SL}(0) & e^{-t \Gamma_L} \tilde{\rho}_{LL}(0) & 0 \\
    0 & 0 & \tilde{\rho}_{00}(t)
  \end{pmatrix}
  ,
\end{equation}
where $\Delta m = m_L - m_S$ (see later), $\Gamma = (\Gamma_S + \Gamma_L)/2$, and because
of~\eqref{eq:tr-rho},
\begin{multline}
  \label{eq:rho00-kaon}
  \tag{\ref{eq:rho-kaon}a}
  \tilde{\rho}_{00}(t) = (1 - e^{-t \Gamma_S}) \tilde{\rho}_{SS}(0) + (1 - e^{-t \Gamma_L}) 
  \tilde{\rho}_{LL}(0) \\
  + 2 \delta_L \Re[(1 - e^{-t (\Gamma + \lambda - i\, \Delta m)}) \tilde{\rho}_{SL}(0)] 
  + \tilde{\rho}_{00}(0).
\end{multline}
%\end{widetext}
Note that, contrary to the case of $\pi^0$, the decoherence parameter
$\lambda$ is no longer irrelevant for evolution of physical states.

The condition that the evolution of the density operator should be
completely positive requires that the inequality
[see~\eqref{eq:cond-app}],
%\begin{multline}
\begin{equation}
  \label{eq:pos-cond}
  \delta_L^2 (1 - 2 e^{-t (\Gamma + \lambda)} \cos(t \Delta m) + e^{-2 t (\Gamma + \lambda)}) %\\
  \leq (1 - e^{-t \Gamma_S}) (1 - e^{-t \Gamma_L}),
\end{equation}
%\end{multline}
must be valid for any $t \geq 0$.  This inequality restricts the range of
the parameters $\lambda$, $\Gamma_S$, $\Gamma_L$, $\Delta m$, and $\delta_L$.  For the physical
values of $\Gamma_S$, $\Gamma_L$, $\Delta m$, and $\delta_L$ for $K^0$ mesons and
corresponding parameters for $B^0$ mesons,
inequality~\eqref{eq:pos-cond} implies an upper bound on $\lambda$ (see
Sec.~\ref{sec:cp-viol-decoh}).

We can write the evolution of the density operator given
by~\eqref{eq:rho-rho} and~\eqref{eq:rho-kaon} in the form of the
operator-sum representation~\eqref{eq:Kraus} with the following Kraus
operators (see Appendix~\ref{sec:oper-sum-repr}):
\begin{subequations}
  \label{eq:kaon-Kraus}
  \allowdisplaybreaks
  \begin{align}
    \hat{E}_0(t) &= \frac{1}{1 - \delta_L^2} \left[e^{-t [(\Gamma_S + \lambda)/2 + i m_S]} 
      \ket{K^0_S} \bra{K^0_S}\right.\notag\\
    &\quad\left. + e^{-t [(\Gamma_L + \lambda)/2 + i m_L]} \ket{K^0_L} \bra{K^0_L}
      - \delta_L \left(e^{-t [(\Gamma_S + \lambda)/2 + i m_S]} \ket{K^0_S} \bra{K^0_L}
      \right.\right.\notag\\
    &\quad\left.\left.
        + e^{-t [(\Gamma_L + \lambda)/2 + i m_L]} \ket{K^0_L} \bra{K^0_S}\right)\right]
    + e^{-t \lambda/2} \ket{0} \bra{0},
  \end{align}
  \begin{align}
    \label{eq:KrausE1}
    \hat{E}_1(t) &= \frac{1}{1 - \delta_L^2} \sqrt{1 - e^{-t \Gamma_S} - \delta_L^2 \frac{\left|1 
          - e^{-t (\Gamma + \lambda - i \Delta m)}\right|^2}{1 - e^{-t \Gamma_L}}} \left(\ket{0} \bra{K^0_S}
      \right.\notag\\
      &\quad\left.- \delta_L \ket{0} \bra{K^0_L}\right),
  \end{align}
  \begin{align}
    \hat{E}_2(t) = \frac{1}{1 - \delta_L^2} \left[\left(\sqrt{1 - e^{-t \Gamma_L}} 
        - \delta_L^2 \frac{1 - e^{-t (\Gamma + \lambda - i \Delta m)}}{\sqrt{1 - e^{-t \Gamma_L}}}\right) 
      \ket{0} \bra{K^0_L}\right.\notag\\
    \quad\left. - \delta_L \left(\sqrt{1 - e^{-t \Gamma_L}} -
        \frac{1 - e^{-t (\Gamma + \lambda - i \Delta m)}}{\sqrt{1 
            - e^{-t \Gamma_L}}}\right) \ket{0} \bra{K^0_S}\right],
    \end{align}
    \begin{gather}
      \hat{E}_3(t) = \frac{e^{-t \Gamma_S/2} \sqrt{1 - e^{-t \lambda}}}{1 - \delta_L^2} 
      \left(\ket{K^0_S} \bra{K^0_S} - \delta_L \ket{K^0_S} \bra{K^0_L}\right),\\
      \hat{E}_4(t) = \frac{e^{-t \Gamma_L/2} \sqrt{1 - e^{-t \lambda}}}{1 - \delta_L^2} 
      \left(\ket{K^0_L} \bra{K^0_L} - \delta_L \ket{K^0_L} \bra{K^0_S}\right),\\
      \hat{E}_5(t) = \sqrt{1 - e^{-t \lambda}} \ket{0} \bra{0}.
    \end{gather}      
  \end{subequations}
Note that \eqref{eq:pos-cond} ensures the reality of the square root
in~\eqref{eq:KrausE1}.

The density operator $\hat{\rho}(t)$ fulfills the master
equation~\eqref{eq:master} with the following Lindblad operators:
\begin{subequations}
  \label{eq:kaon-Lindblad}
  \allowdisplaybreaks
  \begin{equation}
    \hat{L}_1 = \frac{\sqrt{\Gamma_S - \delta_L^2 |\Gamma + \lambda - i\, \Delta m|^2/\Gamma_L}}{1 - \delta_L^2} 
    \left(\ket{0} \bra{K^0_S} - \delta_L \ket{0}{\bra{K^0_L}}\right),\\
  \end{equation}
  \begin{align}
    \hat{L}_2 &= \frac{\sqrt{\Gamma_L} - \delta_L^2 (\Gamma + \lambda - i\, \Delta m)/\sqrt{\Gamma_L}}{1 - \delta_L^2} 
    \ket{0} \bra{K^0_L} \notag\\ &\quad 
    - \delta_L \frac{\sqrt{\Gamma_L} - (\Gamma + \lambda - i\, \Delta m)/\sqrt{\Gamma_L}}{1 - \delta_L^2} 
    \ket{0} \bra{K^0_S},
  \end{align}
  \begin{gather}
    \hat{L}_3 = \frac{\sqrt{\lambda}}{1 - \delta_L^2} \left(\ket{K^0_S} \bra{K^0_S} 
      - \delta_L \ket{K^0_S}{\bra{K^0_L}}\right),\\
    \hat{L}_4 = \frac{\sqrt{\lambda}}{1 - \delta_L^2} \left(\ket{K^0_L} \bra{K^0_L} 
      - \delta_L \ket{K^0_L}{\bra{K^0_S}}\right),\\
    \hat{L}_5 = \sqrt{\lambda} \ket{0} \bra{0},
  \end{gather}
\end{subequations}
and with the Hamiltonian of the form (this is exactly the Hermitian
part of the Weisskopf--Wigner Hamiltonian):
\begin{multline}
  \label{eq:kaon-Hamiltonian}
  \hat{H} = \frac{1}{1 - \delta_L^2} \left\{m_S\ket{K^0_S} \bra{K^0_S}
    + m_L \ket{K^0_L} \bra{K^0_L} \right.\\
    \left.
    - \delta_L \left[(m - i\, \Delta\Gamma/4) \ket{K^0_S} \bra{K^0_L}
      + (m + i\, \Delta\Gamma/4) \ket{K^0_L}
      \bra{K^0_S}\right]\right\},
\end{multline}
where $\Delta\Gamma = \Gamma_S - \Gamma_L$
%The eigenvalues of the Hamiltonian~(\ref{eq:kaon-Hamiltonian}) are 
%\begin{subequations}
%\begin{align}
%  \label{eq:masses}
%  m_L = m + \frac{1}{2} \sqrt{\frac{(\Delta m)^2 + \delta_L^2 (\Delta\Gamma/2)^2}{1 - \delta_L^2}},\\
%  m_S = m - \frac{1}{2} \sqrt{\frac{(\Delta m)^2 + \delta_L^2 (\Delta\Gamma/2)^2}{1 - \delta_L^2}},
%\end{align}
%\end{subequations}
and $m = (m_L + m_S)/2$ is the mean $K^0$ mass (measured
experimentally; $m = \unit{497.648}{\mega\electronvolt}/c^2$ for
$K^0$, $m_{B^0} = \unit{5279.4}{\mega\electronvolt}/c^2$ for $B^0$
\cite{eidelman04:abb}):
\begin{subequations}
\begin{align}
  \allowdisplaybreaks
  m_{K^0} &= \bra{K^0}\hat{H}\ket{K^0} = m,\\
  m_{\bar{K}^0} &= \bra{\bar{K}^0}\hat{H}\ket{\bar{K}^0} = m.
\end{align}
\end{subequations}
Note that $m_{K^0} = m_{\bar{K}^0}$, as it is required by $CPT$
theorem.  $\Delta m$ is measured by an observation of $K^0$ flavor
oscillation.  This finally gives us the interpretation of $m_S$ and
$m_L$, which appeared in~(\ref{eq:rho-kaon}), as the expectation
values of Hamiltonian in $\ket{K^0_L}$ and $\ket{K^0_S}$ states:
\begin{subequations}
  \label{eq:mSmL}
  \begin{align}
    \allowdisplaybreaks
    m_L &= \bra{K^0_L}\hat{H}\ket{K^0_L} = m + \Delta m/2,\\
    m_S &= \bra{K^0_S}\hat{H}\ket{K^0_S} = m - \Delta m/2.
  \end{align}
\end{subequations}
We would like to stress that the $\ket{K^0_S}$ and $\ket{K^0_L}$
\emph{are not eigenstates} of the
Hamiltonian~(\ref{eq:kaon-Hamiltonian}).

%\section{Probabilities, averages, and state reductions}
%\label{sec:prob-aver-state}

%In this section we show that some issues that have no unambiguous
%answers in Weisskopf--Wigner approach can be solved by means of the
%presented approach.

From~\eqref{eq:rho-kaon}, it follows that the probabilities of
detecting $K^0$ and $\bar{K}^0$ are given by
\begin{subequations}
  \allowdisplaybreaks
  \label{eq:probabilities-K}
  \begin{align}
    p_{K^0}(t) &= \tr[\hat{\rho}(t) \ket{K^0} \bra{K^0}] 
    = \frac{1 + \delta_L}{2} \left\{e^{-t \Gamma_S} \tilde{\rho}_{SS}(0) 
      + e^{-t \Gamma_L} \tilde{\rho}_{LL}(0) \right.\notag\\ &\quad\left.
      + 2 \Re\left[e^{-t (\Gamma + \lambda - i \Delta m)} \tilde{\rho}_{SL}(0)\right]\right\},
    \label{eq:prob-K}
  \end{align}
  \begin{align}
    p_{\bar{K}^0}(t) &= \tr[\hat{\rho}(t) \ket{\bar{K}^0} \bra{\bar{K}^0}]
    = \frac{1 - \delta_L}{2} \left\{e^{-t \Gamma_S} \tilde{\rho}_{SS}(0) 
      + e^{-t \Gamma_L} \tilde{\rho}_{LL}(0) \right.\notag\\ &\quad\left.
     - 2 \Re\left[e^{-t (\Gamma + \lambda - i \Delta m)} \tilde{\rho}_{SL}(0)\right]\right\}.
   \label{eq:prob-antyK}
  \end{align}
\end{subequations}
If the initial state is $\ket{K^0}$, these probabilities are
\begin{subequations}
  \begin{gather}
    \allowdisplaybreaks
    p_{K^0}(t)
    = \frac{1}{4} \left[e^{-t \Gamma_S} + e^{-t \Gamma_L} 
      + 2 e^{-t (\Gamma + \lambda)} \cos(t\, \Delta m)\right],\\
    p_{\bar{K}^0}(t)
    = \frac{1}{4} \frac{1 - \delta_L}{1 + \delta_L} 
    \left[e^{-t \Gamma_S} + e^{-t \Gamma_L} - 2 e^{-t (\Gamma + \lambda)} \cos(t\, \Delta m)\right].
  \end{gather}
\end{subequations}

%% Note that for $B^0$ mesons we have $\Gamma_{B^0_H} \approx \Gamma_{B^0_L}$.  If the
%% initial state is $\ket{B^0}$, the probabilities are
%% \begin{subequations}
%%   \label{eq:probabilities-B}
%%   \begin{align}
%%     p_{B^0}(t) %&= \tr[\hat{\rho}(t) \ket{B^0} \bra{B^0}]\notag\\
%% %    &= \frac{e^{-t \Gamma}}{1 + |\epsilon_{B^0}|^2} \left[\tilde{\rho}_{LL}(0) + |\epsilon_{B^0}|^2 \tilde{\rho}_{HH}(0)
%% %      + 2\Re\left(\epsilon_{B^0}^* e^{-t (\lambda - i \Delta m_{B^0})} \tilde{\rho}_{LH}(0)\right)\right]\\
%%     \label{eq:prob-B}
%%     &= \frac{e^{-t \Gamma}}{2} \left[1 + e^{-t \lambda} \cos(t \Delta m_{B^0})\right],\\
%%     p_{\bar{B}^0}(t) %&= \tr[\hat{\rho}(t) \ket{\bar{B}^0} \bra{\bar{B}^0}]\notag\\
%%     \label{eq:prob-antyB}
%%     &= \frac{e^{-t \Gamma}}{2} \frac{|1 - \epsilon_{B^0}|^2}{|1 + \epsilon_{B^0}|^2} 
%%     \left[1 - e^{-t \lambda} \cos(t \Delta m_{B^0})\right].
%%   \end{align}
%% \end{subequations}
%% %\end{widetext}

The strangeness operator for $K^0$ is $\hat{S} = \ket{K^0} \bra{K^0} -
\ket{\bar{K}^0} \bra{\bar{K}^0}$, and its average is
%\begin{multline}
\begin{align}
  \langle\hat{S}\rangle &= \tr[\hat{\rho}(t) \hat{S}] 
  = \delta_L \left[e^{-t \Gamma_S} \tilde{\rho}_{SS}(0) 
    + e^{-t \Gamma_L} \tilde{\rho}_{LL}(0)\right] \notag\\ &\quad
  + 2 \Re\left[e^{-t (\Gamma + \lambda - i\, \Delta m)} \tilde{\rho}_{SL}(0)\right].
  \label{eq:strangeness}
\end{align}
%\end{multline}
Regardless of the initial state $\lim_{t \to \infty} \langle\hat{S}\rangle = 0$, as
expected, because in the limit $t \to \infty$ we have vacuum only.  If the
initial state is $\ket{K^0}$, we get
\begin{equation}
  \label{eq:S-average}
  \langle\hat{S}\rangle = \frac{1}{1 + \delta_L} \left(e^{-t (\Gamma + \lambda)} \cos(t\, \Delta m) 
    + \frac{\delta_L}{2} (e^{-t \Gamma_S} + e^{-t \Gamma_L})\right).
\end{equation}

Finally, we would like to point out that we are dealing with the
Hermitian Hamiltonian and unit-trace density operator.  This assures
us that there is no ambiguity in calculating either conditional or
joint probabilities for the results of measurements performed on the
system.  Moreover, we can unambiguously determine the states after the
projective measurement using the standard quantum mechanical
procedures (i.e., the von Neumann postulate of the state reduction).

\section{CP violation and decoherence}
\label{sec:cp-viol-decoh}

\begin{figure}
  \centering
  \includegraphics[width=\columnwidth]{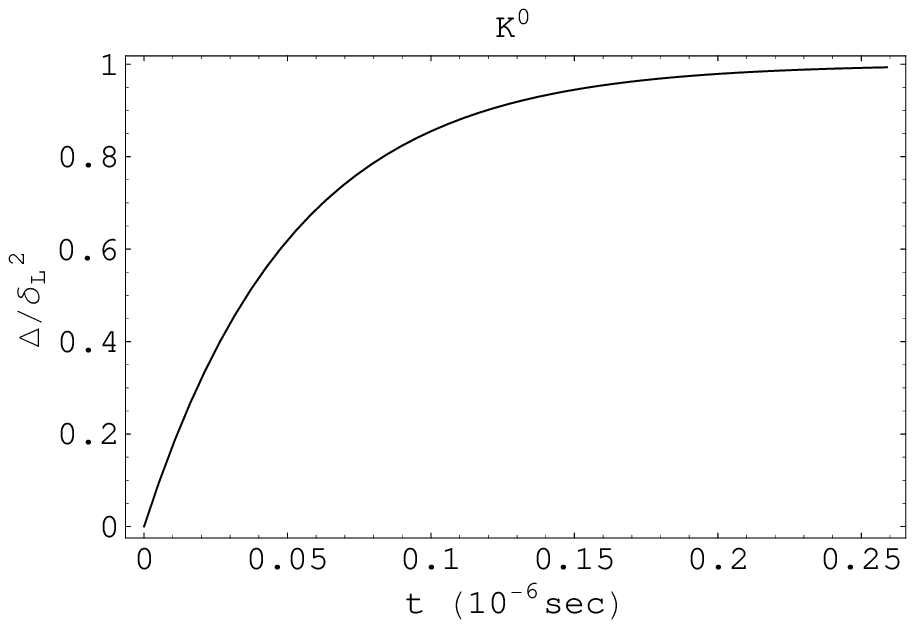}
  \includegraphics[width=\columnwidth]{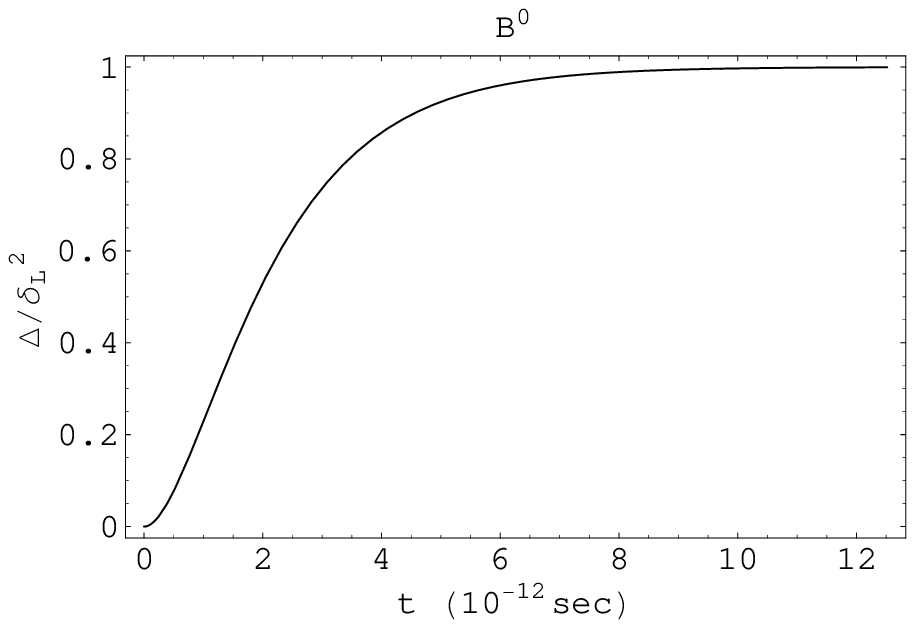}
  \caption{The discriminant of inequality~\eqref{eq:pos-cond} as a
    function of time for $K^0$ and $B^0$; in both cases $\Delta \geq 0$,
    implying the validity of~\eqref{eq:condition}.}
  \label{fig:delta}
\end{figure}

Now, let us analyze the inequality~\eqref{eq:pos-cond} in more detail.
Taking into account that $\lambda \geq 0$, we can treat~\eqref{eq:pos-cond} as a
quadratic inequality in $e^{-t (\Gamma + \lambda)}$.  This inequality has real
solutions provided that its discriminant $\Delta$ fulfills
\begin{equation}
  \label{eq:condition}
  \Delta/\delta_L^2 = [(1 - e^{-t \Gamma_S}) (1 - e^{-t \Gamma_L}) - \delta_L^2 \sin^2(t\, \Delta m)] \geq 0
\end{equation}
for any $t \geq 0$, where, according to \cite{eidelman04:abb}: $\Delta m =
\unit{0.5292 \times 10^{10}}{\reciprocal\second}$, $\tau_S \equiv 1/\Gamma_S =
\unit{0.8953 \times 10^{-10}}{\second}$, $\tau_L \equiv 1/\Gamma_L = \unit{5.18 \times
  10^{-8}}{\second}$.  $\Delta m_{B^0} = \unit{0.502 \times
  10^{12}}{\reciprocal\second}$ ($\Delta m_{B^0} \approx m_{B^0_H} - m_{B^0_L}$),
$\tau \equiv 1/\Gamma = \unit{1.536 \times 10^{-12}}{\second}$, and $\Re(\epsilon_{B^0})/(1 +
|\epsilon_{B^0}|^2) = 0.5 \times 10^{-3}$.  Fortunately, the
inequality~\eqref{eq:condition} holds for any $t \geq 0 $ for both $K^0$
and $B^0$ mesons because the first term rapidly grows from $0$ to $1$
while the second one oscillates between $0$ and $\delta_L^2 \sim 10^{-6}$ (see
Fig.~\ref{fig:delta}).
Indeed, the series expansion of~\eqref{eq:condition} is
\begin{equation}
  \label{eq:cond-series}
  \Delta/\delta_L^2 \simeq [\Gamma_S \Gamma_L - \delta_L^2 (\Delta m)^2] t^2 + O(t^3).
\end{equation}
Therefore the condition
\begin{equation}
  \label{eq:cond-delta}
  \delta_L \leq \frac{\sqrt{\Gamma_S \Gamma_L}}{\Delta m}
\end{equation}
is necessary for the existence of the solutions of
inequality~\eqref{eq:pos-cond}.  This condition is satisfied for both
$K^0$ and $B^0$ mesons.

\begin{figure}
  \centering
  \includegraphics[width=\columnwidth]{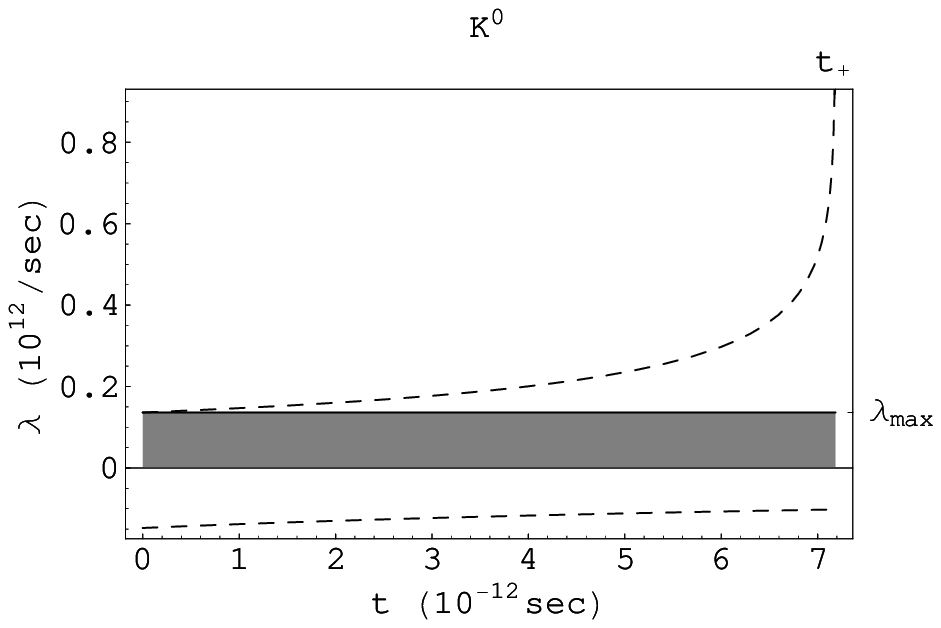}
  \includegraphics[width=\columnwidth]{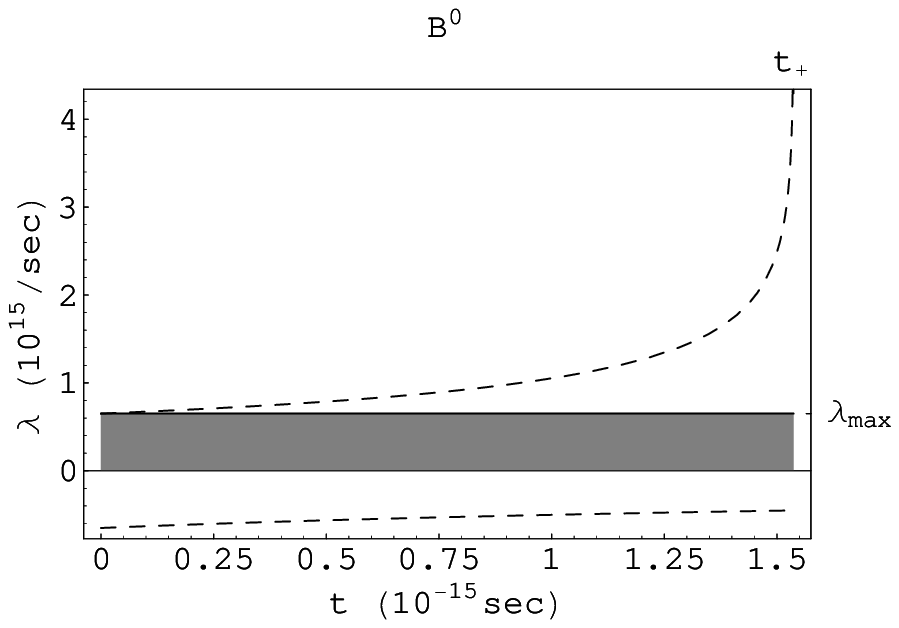}
  \caption{The allowed region for the decoherence parameter (gray) for
    $K^0$ and $B^0$ mesons: dashed curves are the upper and lower
    bound from~\eqref{eq:double-log-ineq}; the lower bound is always
    negative.  The values of $t_+$ and $\lambda_{\mathrm{max}}$ for $K^0$
    and $B^0$ are given in the text.}
  \label{fig:inequality}
\end{figure}

Now, let us analyze the restrictions imposed by the
inequality~\eqref{eq:pos-cond} on the range of the decoherence
parameter $\lambda$.  From~\eqref{eq:pos-cond} we have, for any $t \geq 0$,
\begin{equation}
  \label{eq:double-ineq}
  \cos(t\, \Delta m) - \frac{\sqrt{\Delta}}{\delta_L^2}  \leq e^{-t (\Gamma + \lambda)}  
  \leq \cos(t\, \Delta m) + \frac{\sqrt{\Delta}}{\delta_L^2}.
\end{equation}
The left inequality gives a restriction only when its left-hand side
is positive, which holds for $K^0$ and $B^0$ only for $t \leq t_+$, where
$t_+ \approx \unit{7.18517 \times 10^{-12}}{\second}$ for $K^0$ and $t_+ \approx
\unit{1.53677 \times 10^{-15}}{\second}$ for $B^0$.  Therefore
%\begin{widetext}
\begin{subequations}
  \label{eq:double-log-ineq}
  \begin{gather}
    \lambda \leq -\frac{1}{t}\ln[\cos(t\, \Delta m) - \sqrt{\Delta}] - \Gamma,\\
    \lambda \geq -\frac{1}{t}\ln[\cos(t\, \Delta m) + \sqrt{\Delta}] - \Gamma,
  \end{gather}
\end{subequations}
%\end{widetext}
where the upper inequality must hold for $0 \leq t \leq t_+$ and the lower
one for any $t \geq 0$; moreover, as mentioned earlier, $\lambda \geq 0$.  These
bounds are presented in Fig.~\ref{fig:inequality} for $K^0$ and $B^0$.
Thus, for $K^0$ and $B^0$,
\begin{equation}
  \label{eq:lambda-bound}
  \begin{split}
    0 \leq \lambda \leq \lambda_{\mathrm{max}} &= \inf_{0 \leq t \leq t_+} \left(-\frac{1}{t}\ln[\cos(t\, \Delta m) 
      - \sqrt{\Delta}] - \Gamma\right) \\
    & \simeq \frac{1}{\delta_L} \sqrt{\Gamma_S \Gamma_L - \delta_L^2 (\Delta m)^2} - \Gamma,
  \end{split}
\end{equation}
where the last equality comes from the first-order expansion.  For
$K^0$ we get $\lambda_{\mathrm{max}} = \unit{1.3629 \times
  10^{11}}{\reciprocal\second}$ and for $B^0$ we get $\lambda_{\mathrm{max}}
= \unit{6.5039 \times 10^{14}}{\reciprocal\second}$.  The experimental
value for the decoherence parameter in an entangled $K^0\bar{K}^0$ system is
$\lambda = \unit{(1.84^{+2.50}_{-2.17}) \times 10^{-12}}{\mega\electronvolt} =
\unit{2.80^{+3.80}_{-3.30} \times 10^9}{\reciprocal\second}$
\cite{apostolakis98:abb} and for a $B^0\bar{B}^0$ system is $\lambda =
\unit{(-47 \pm 76) \times 10^{-12}}{\mega\electronvolt} = \unit{(-0.71 \pm
  1.15) \times 10^{11}}{\reciprocal\second}$ \cite{bertlmann01b} and they
fit in the allowed range.

\section{Conclusions}
\label{sec:conclusions}

In this paper, we have found the probability-preserving description of
the decaying particle within the framework of quantum mechanics of
open systems, taking into account the superselection rule prohibiting
the superposition of the particle and vacuum.  It has been shown that
some limitations of the Weisskopf--Wigner approach can be removed if
we assume that the particle can be found in one of its possible states
as well as in the state of \emph{the absence of the particle}, i.e.,
in the \emph{vacuum} state.  In our approach the evolution of the
system is given by a family of completely positive trace-preserving
maps forming a one-parameter dynamical semigroup; thus the
\emph{Hamiltonian is Hermitian} and therefore the \emph{reduced
  density operator has a unit trace}.  It should be noted that we have
used the dynamical semigroup approach for the description of the
\emph{entire unstable particle}, not only for the description of its
decoherence, as in
\cite{benatti96,benatti97a,benatti97b,benatti98,quant-ph/0410028}.
The advantage of the introduced approach is that there is no ambiguity
in calculating either conditional or joint probabilities for the
results of measurements performed on the system.  Furthermore, we can
unambiguously determine the states after the measurement using the
standard quantum mechanical procedures (i.e., the von Neumann
postulate on the state reduction).  We have also shown that there are
restrictions on the possible strength of decoherence that arise as a
remarkable side effect of completely positive evolution of the
particle.  Moreover, we have found the operator sum representation
(Kraus representation) for the general evolution of such systems,
which allows one to write the evolution for systems with two or more
particles.  This is extremely useful if we study quantum correlations
between unstable particles.  Moreover, we have shown that the decay of
the particle can be regarded as a Markov process by finding explicitly
the Lindblad form of the master equation for such a system.

\begin{acknowledgments}
  We would like to acknowledge fruitfull discussions with R.  Alicki,
  J. Ciborowski, J. K{\l}osi{\'n}ski, A. Kossakowski, and M.
  W{\l}odarczyk.  This work is supported by the Polish Ministry of
  Scientific Research and Information Technology under Grant
  No.~PBZ/MIN/008/P03/2003 and by the University of Lodz.
\end{acknowledgments}

\appendix
\boldmath
\section{Time dependence of density matrices for $K^0$}
\label{sec:time-depend}
\unboldmath

In this appendix we show that the evolution given
by~\eqref{eq:rho-kaon} is the most general completely positive
trace-preserving linear map that possesses the semigroup property and
leads to the Weisskopff--Wigner evolution~\eqref{eq:WW-K}.  This
implies that we have well-defined $\tilde{\rho}_{SS}(t) =
\tilde{\rho}_{SS}(0) e^{-t \Gamma_S}$ and $\tilde{\rho}_{LL}(t) =
\tilde{\rho}_{LL}(0) e^{-t \Gamma_L}$.  Because kaons carry some quantum
numbers like strangeness, we impose the superselection rule from the
very beginning.

The most general form of the evolution can be written in the form
\begin{widetext}
\begin{equation}
  \label{eq:S-rho:noCP}
  \tilde{\mathcal{S}}_t \tilde{\rho}(0) =
  \begin{pmatrix}
    e^{-t \Gamma_S} \tilde{\rho}_{SS}(0) & \sum\limits_{i,j=S,L,0} A_{ij}(t) \tilde{\rho}_{ij}(0) &
    \sum\limits_{i,j=S,L,0} B_{ij}(t) \tilde{\rho}_{ij}(t) \\
    \sum\limits_{i,j=S,L,0} A_{ij}^*(t) \tilde{\rho}_{ij}^*(0) & e^{-t \Gamma_L} \tilde{\rho}_{LL}(0) &
    \sum\limits_{i,j=S,L,0} C_{ij}(t) \tilde{\rho}_{ij}(0) \\
    \sum\limits_{i,j=S,L,0} B_{ij}^*(t) \tilde{\rho}_{ij}^*(0) &
    \sum\limits_{i,j=S,L,0} C_{ij}^*(t) \tilde{\rho}_{ij}^*(0) &
    \sum\limits_{i,j=S,L,0} D_{ij}(t) \tilde{\rho}_{ij}(0)
  \end{pmatrix}
  .
\end{equation}
\end{widetext}
The superselection rule causes that the only nonvanishing $B$'s and
$C$'s are $B_{i0}(t)$, $B_{0i}(t)$, $C_{i0}(t)$, and $C_{0i}(t)$;
moreover, since $\tilde{\rho}_{00}(t)$ must be real, $D_{LS}(t) =
D_{SL}^*(t)$, $D_{0S}(t) = D_{S0}^*(t)$, $D_{0L}(t) = D_{L0}^*(t)$,
and the other $D$'s are real functions.  The initial conditions are
$A_{SL}(0) = B_{S0}(0) = C_{L0}(0) = 1$ and the other functions vanish
at $t = 0$; moreover, $D_{00}(t) \equiv 1$ because the vacuum must be a
fixed point of this dynamics, i.e.,
\begin{equation}
  \label{eq:vacuum-evolution}
  \rho(0) = \ket{0} \bra{0} \Rightarrow \forall t \geq 0\colon \rho(t) = \ket{0} \bra{0}.
\end{equation}

The corresponding Choi's matrix is
\begin{widetext}
\begin{equation}
  \label{eq:Choi-K-noCP}
  \Choi\tilde{\mathcal{S}}_t =
  \begin{pmatrix}
    e^{-t \Gamma_S} & A_{SS}(t) & 0 & 0 & A_{SL}(t) & 0 & 0 & A_{S0}(t) & B_{S0}(t) \\
    A_{SS}^*(t) & 0 & 0 & A_{LS}^*(t) & 0 & 0 & A_{0S}^*(t) & 0 & C_{S0}(t) \\
    0 & 0 & D_{SS}(t) & 0 & 0 & D_{SL}(t) & B_{0S}^*(t) & C_{0S}^*(t) & D_{S0}(t) \\
    0 & A_{LS}(t) & 0 & 0 & A_{LL}(t) & 0 & 0 & A_{L0}(t) & B_{L0}(t) \\
    A_{SL}^*(t) & 0 & 0 & A_{LL}^*(t) & e^{-t \Gamma_L} & 0 & A_{0L}^*(t) & 0 & C_{L0}(t) \\
    0 & 0 & D_{SL}^*(t) & 0 & 0 & D_{LL}(t) & B_{0L}^*(t) & C_{0L}^*(t) & D_{L0}(t) \\
    0 & A_{0S}(t) & B_{0S}(t) & 0 & A_{0S}(t) & B_{0L}(t) & 0 & A_{00}(t) & B_{00}(t) \\
    A_{S0}^*(t) & 0 & C_{0S}(t) & A_{L0}^*(t) & 0 & C_{0L}(t) & A_{00}^*(t) & 0 & C_{00}(t) \\
    B_{S0}^*(t) & C_{S0}^*(t) & D_{S0}^*(t) & B_{L0}^*(t) & C_{L0}^*(t) & D_{L0}^*(t) & B_{00}^*(t) & C_{00}^*(t) & 1
  \end{pmatrix}
\end{equation}
\end{widetext}
The positivity of this matrix requires that from the functions $A$'s,
$B$'s, and $C$'s the only nonvanishing functions are $A_{SL}(t)$, $B_{S0}(t)$,
and $C_{L0}(t)$.  Moreover, the condition on the trace of $\hat{\rho}(t)$
implies that $\tr[\rho(t)] = \tr[V \tilde{\rho}(t) V^{\dag}] = 1$, so we have
%\begin{widetext}
\begin{multline}
  \label{eq:tr-cond}
  [e^{-t \Gamma_S} + D_{SS}(t)] \tilde{\rho}_{SS}(0) + [e^{-t \Gamma_L} +
  D_{LL}(t)] \tilde{\rho}_{LL}(0)\\
  + 2 \Re\{[\delta_L A_{SL}(t) + D_{SL}(t)] \tilde{\rho}_{SL}(0)\} \\
  + 2\Re[D_{S0}(t) \tilde{\rho}_{S0}(0)] + 2\Re[D_{L0}(t) \tilde{\rho}_{L0}(0)]\\
  = \tilde{\rho}_{SS}(0) + \tilde{\rho}_{LL}(0) + 2 \delta_L \Re[\tilde{\rho}_{SL}(0)].
\end{multline}
%\end{widetext}
Because this equation must be valid for any $\hat{\rho}(0)$, this implies
that $D_{i0}(t) = 0$ for $i \neq 0$, and
\begin{subequations}
  \label{eq:00:K-noCP}
  \begin{gather}
    \label{eq:D00}
    D_{00}(t) \equiv 1,\\
    \label{eq:DSS}
    D_{SS}(t) = 1 - e^{-t \Gamma_S},\\
    \label{eq:DLL}
    D_{LL}(t) = 1 - e^{-t \Gamma_L},\\
    \label{eq:Doff}
    D_{SL}(t) = \delta_L [1 - A_{SL}(t)].
  \end{gather}
\end{subequations}

Therefore, the matrix~\eqref{eq:Choi-K-noCP} is positive, provided that
\begin{subequations}
  \allowdisplaybreaks
  \label{eq:positive:K}
  \begin{gather}
    \label{eq:positive:K12}
    |A_{SL}(t)|^2 \leq e^{-t (\Gamma_S + \Gamma_L)},\\
    \label{eq:positive:K13}
    |B_{S0}(t)|^2 \leq e^{-t \Gamma_L},\\
    \label{eq:positive:K23}
    |C_{L0}(t)|^2 \leq e^{-t \Gamma_S},
  \end{gather}
\begin{multline}
%\begin{equation}
    \label{eq:positive:K-mix}
    |B_{S0}(t)|^2 e^{-t \Gamma_L} + |C_{L0}(t)|^2 e^{-t \Gamma_S} + |A_{SL}(t)|^2 \\
    - 2\Re[A_{SL}(t) B_{S0}^*(t) C_{L0}(t)]
    \leq e^{-t (\Gamma_S + \Gamma_L)} ,
%\end{equation}
\end{multline}
  and
  \begin{equation}
    \label{eq:positive-Doff}
    |D_{SL}(t)|^2 \leq D_{SS}(t) D_{LL}(t).
  \end{equation}
\end{subequations}

Taking into account the composition law for the $\mathcal{S}_t$
superoperator and \eqref{eq:positive:K12}--\eqref{eq:positive:K23}, we
get
\begin{subequations}
  \label{eq:off-diag:K}
  \begin{gather}
    \label{eq:off-diag:K12}
    A_{SL}(t) = e^{-t \left[(\Gamma_S + \Gamma_L)/2 + a - i \mu_{SL}\right]} ,\\
    B_{S0}(t) = e^{-t \left[\Gamma_S/2 + b + i \mu_{S}\right]},\\
    C_{L0}(t) = e^{-t \left[\Gamma_L/2 + c + i \mu_{L}\right]},
  \end{gather}
\end{subequations}
where $a, b, c, \mu_{SL}, \mu_{S}, \mu_{L} \geq 0$.  Now, one can observe that
the concrete time dependence of the functions $B_{S0}(t)$ and
$C_{L0}(t)$ is irrelevant, since the superselection rule requires that
$\tilde{\rho}_{S0}(0) = \tilde{\rho}_{L0}(0) = 0$ and, therefore,
$\tilde{\rho}_{S0}(t)$ and $\tilde{\rho}_{L0}(t)$ still remain vanishing (as
required), so, without loss of any generality, we can choose $a = b =
c = \lambda$ and $\mu_{S} = m_S$, $\mu_{L} = m_L$, and then $\mu_{SL} = \Delta m$.
Thus, finally, we can put
\begin{subequations}
  \label{eq:off-diag:K-noCP}
  \begin{gather}
    A_{SL}(t) = e^{-t \left[(\Gamma_S + \Gamma_L)/2 + \lambda - i\, \Delta m\right]} ,\\
    B_{S0}(t) = e^{-t \left[\Gamma_S/2 + \lambda + i m_S\right]},\\
    C_{L0}(t) = e^{-t \left[\Gamma_L/2 + \lambda + i m_L\right]},
    \intertext{and, consequently,}
    D_{SL}(t) = \delta_L \left(1 - e^{-t \left[(\Gamma_S + \Gamma_L)/2 + \lambda - i\, \Delta m\right]}\right).
  \end{gather}
\end{subequations}

Now, taking the solutions \eqref{eq:D00}--\eqref{eq:DLL}
and~\eqref{eq:off-diag:K-noCP}, one can easily see that among
conditions~\eqref{eq:positive:K} for positivity of the matrix
$\Choi\tilde{\mathcal{S}}_t$ the only one that is not identically
fulfilled is~\eqref{eq:positive-Doff}; it can be written in the form
\begin{equation}
  \label{eq:cond-app}
%   \forall t\geq 0\colon 
   \delta_L^2 \left|1 - e^{-t [(\Gamma_S + \Gamma_L)/2 + \lambda - i\, \Delta m]}\right|^2 
   \leq (1 - e^{-t \Gamma_S}) (1 - e^{-t \Gamma_L}),
\end{equation}
for any $t \geq 0$.

\boldmath
\section{Operator sum representation for evolution of $K^0$}
\label{sec:oper-sum-repr}
\unboldmath

Let us denote
\begin{equation}
  \label{eq:g}
  g \equiv V^{\dag}{} V =
  \begin{pmatrix}
    1 & \delta_L & 0 \\
    \delta_L & 1 & 0 \\
    0 & 0 & 1
  \end{pmatrix}
  .
\end{equation}
We have
\begin{equation}
  \label{eq:1}
  \tr[\rho(t)] = \tr[\tilde{\rho}(t) g],
\end{equation}
and therefore we can easily find Eq.~\eqref{eq:tr-rho}.

Now, let us find the operator sum representation for the evolution
given by the map~\eqref{eq:rho-kaon}.  Let us define the set of
matrices $\breve{E}_i(t)$, such that
\begin{equation}
  \label{eq:Kraus-2bases}
  \tilde{\rho}(t) = \sum_i \breve{E}_i(t) \tilde{\rho}(0) \breve{E}_i^{\dag}(t).
\end{equation}
%so
%\begin{equation}
%  \label{eq:E-E}
%  E_i(t) = V \tilde{E}_i(t) V^{-1}.
%\end{equation}
%The normalization condition $\sum_i E_i^{\dag}(t) E_i(t) = I$ rewritten in
%the~\eqref{eq:H-basis} reads
%\begin{equation}
%  \label{eq:E-norm}
%  \sum_i \tilde{E}_i^{\dag}(t) g \tilde{E}_i(t) = g.
%\end{equation}
After a little algebra, we find that the matrices $\breve{E}_i(t)$ are
\begin{subequations}
  \allowdisplaybreaks
  \label{eq:Kraus-noCP}
  \begin{gather}
    \breve{E}_0(t) =
    \begin{pmatrix}
      e^{-t \left[(\Gamma_S + \lambda)/2 + i m_S\right]} & 0 & 0 \\
      0 & e^{-t \left[(\Gamma_L + \lambda)/2 + i m_L\right]} & 0 \\
      0 & 0 & e^{-t \lambda/2} 
    \end{pmatrix},\\
    \breve{E}_1(t) = \sqrt{1 - e^{-t \Gamma_S} 
      - \delta_L^2 \frac{\left|1 - e^{-t (\Gamma + \lambda - i \Delta m)}\right|^2}{1 - e^{-t \Gamma_L}}}
    \begin{pmatrix}
      0 & 0 & 0 \\
      0 & 0 & 0 \\
      1 & 0 & 0
    \end{pmatrix}
    ,\\
    \breve{E}_2(t) = \sqrt{1 - e^{-t \Gamma_L}}
    \begin{pmatrix}
      0 & 0 & 0 \\
      0 & 0 & 0 \\
      \delta_L \frac{1 - e^{-t (\Gamma + \lambda - i \Delta m)}}{1 - e^{-t \Gamma_L}} 
      & 1 & 0
    \end{pmatrix}
    ,\\
    \breve{E}_3(t) = e^{-t \Gamma_S/2} \sqrt{1 - e^{-t \lambda}}
    \begin{pmatrix}
      1 & 0 & 0 \\
      0 & 0 & 0 \\
      0 & 0 & 0
    \end{pmatrix},\\
    \breve{E}_4(t) = e^{-t \Gamma_L/2} \sqrt{1 - e^{-t \lambda}}
    \begin{pmatrix}
      0 & 0 & 0 \\
      0 & 1 & 0 \\
      0 & 0 & 0
    \end{pmatrix},\\
    \breve{E}_5(t) = \sqrt{1 - e^{-t \lambda}}
    \begin{pmatrix}
      0 & 0 & 0 \\
      0 & 0 & 0 \\
      0 & 0 & 1
    \end{pmatrix}.
  \end{gather}
\end{subequations}

These matrices can be helpful in finding the Kraus operators $\hat{E}_i(t)$.
Indeed, 
\begin{equation}
  \label{eq:2}
  \sum_i V \breve{E}_i(t) \tilde{\rho}(0) \breve{E}_i^{\dag}(t) V^{\dag}{} = \rho(t) 
  = \sum_i E_i(t) \rho(0) E_i^{\dag}(t),
\end{equation}
where matrices on the right-hand side are written in the orthonormal
basis~\eqref{eq:CP-basis}.  This gives
\begin{equation}
  \label{eq:E-E:1}
  E_i(t) = V \breve{E}_i(t) V^{-1}.
\end{equation}

More interesting than finding explicitly the matrices $E_i(t)$'s is
finding the decomposition of $\hat{E}_i(t)$ into the sum
%\begin{widetext}
\begin{align}
  \allowdisplaybreaks
  \hat{E}_i(t) &= 
  \tilde{E}_i(t)_{SS} \ket{K^0_S} \bra{K^0_S} + \tilde{E}_i(t)_{SL} \ket{K^0_S} \bra{K^0_L} 
  + \tilde{E}_i(t)_{S0} \ket{K^0_S} \bra{0} \notag\\ &\quad
  + \tilde{E}_i(t)_{LS} \ket{K^0_L} \bra{K^0_S} 
  + \tilde{E}_i(t)_{LL} \ket{K^0_L} \bra{K^0_L} + \tilde{E}_i(t)_{L0} \ket{K^0_L} \bra{0}
  \notag\\ &\quad
  + \tilde{E}_i(t)_{0S} \ket{0} \bra{K^0_S} + \tilde{E}_i(t)_{0L} \ket{0} \bra{K^0_L}
  + \tilde{E}_i(t)_{00} \ket{0} \bra{0}.
  \label{eq:3}
\end{align}
%\end{widetext}
Using Eqs.~\eqref{eq:H-basis} and~\eqref{eq:E-E:1}, we get finally that
\begin{equation}
  \label{eq:E-E:2}
  \tilde{E}_i(t) = V^{-1} E_i(t) {V^{-1}}^{\dag}{} = \breve{E}_i(t) g^{-1}.
\end{equation}
%\bibliography{pdg,quant-ph}

\end{document}